# Implementation of Quantum and Classical Discrete Fractional Fourier Transforms

Steffen Weimann,[1*] Armando Perez-Leija,[1*] Maxime Lebugle,[1*] Robert Keil,[2] Malte Tichy,[3] Markus Gräfe,[1] Rene Heilmann,[1] Stefan Nolte,[1] Hector Moya-Cessa,[4] Gregor Weihs,[2] Demetrios N. Christodoulides,[5] and Alexander Szameit[1†]

[1]*Institute of Applied Physics, Abbe School of Photonics, Friedrich-Schiller-Universität Jena, Max-Wien Platz 1, 07743 Jena, Germany*
[2]*Institut für Experimentalphysik, Universität Innsbruck, Technikerstraße 25, Innsbruck 6020, Austria*
[3]*Department of Physics and Astronomy, University of Aarhus, 8000 Aarhus, Denmark*
[4]*INAOE, Coordinacion de Optica, Luis Enrique Erro No. 1, Tonantzintla, Puebla 72840, Mexico*
[5]*CREOL, The College of Optics & Photonics, University of Central Florida, Orlando, Florida 32816, USA*
* these authors contributed equally
† alexander.szameit@uni-jena.de

**Fourier transforms are ubiquitous mathematical tools in basic and applied sciences. We here report classical and quantum optical realizations of the discrete fractional Fourier transform, a generalization of the Fourier transform. In the integrated configuration used in our experiments, the order of the transform is mapped onto the longitudinal coordinate, thus opening up the prospect of simultaneously observing all transformation orders. In the context of classical optics, we implement discrete fractional Fourier transforms, both integer and fractional, of exemplary wave functions and experimentally demonstrate the shift theorem. Moreover, we apply this approach in the quantum realm to transform separable and highly entangled biphoton wave functions. The proposed approach is versatile and could find applications in various fields where Fourier transforms are essential tools, such as quantum chemistry and biology, physics and mathematics.**

Two hundred years ago, Joseph Fourier introduced a major concept in mathematics, the so-called Fourier Transform (FT). It was not until 1965, when Cooley and Tukey developed the "Fast Fourier Transform" algorithm (FFT), that Fourier analysis became a standard tool in contemporary sciences (1). Two crucial requirements in this algorithm are the discretization and truncation of the domain where the signals to be transformed are defined. These requirements are always satisfiable since observable quantities in physics must be well-behaved and finite in extension and magnitude.

An outstanding application of the FFT was put forward by Shor that could provide an exponential speedup in factoring integer numbers (2) by exploiting the principles of quantum mechanics. Following this development, a plethora of theoretical and experimental work appeared aiming to describe how discrete Fourier transforms (DFTs) can be realized and applied in various and diverse settings (3). In parallel, Namias made another significant leap with the introduction of the fractional Fourier transform (FrFT), which contains the FT as a special case (4). Several investigations quickly followed, leading to a more general theory of joint time-frequency signal representations and demonstrating the potential of the FrFT in areas such as wave propagation, signal processing and differential equations (3, 5, 6). Subsequently, the need for discretization of this generalized FT led to the introduction of the Discrete Fractional Fourier Transform (DFrFT) operating on a finite grid in a way similar to that of a discrete FT (7). Along those lines, several versions of the DFrFT have been introduced (3). In this work we focus on the experimental demonstration of an optical DFrFT, which can be obtained after a rigorous discretization of the continuous FrFT. This transformation is based on the so-called Fourier-Kravchuk transform (7) and can be equally applied to classical and quantum states. Throughout our paper we simply refer to this transform as DFrFT.

We foresee that the inherent versatility of our approach will open the door to many interesting applications in several branches of science. In this regard one may mention the determination of quasiprobability distributions in quantum mechanics (8), analysis of wave fields in quantum optics (9), the development of swept-frequency and time-variant filters, spatial multiplexing in signal processing (10, 11), pattern recognition and generalized space-frequency distributions (12), the implementation of the Radon and wavelet transforms (13), phase retrieval algorithms in quantum chemistry (14-17), signal detection, radar analysis, tomography, data compression, (3) etc.

Similarly to its continuous counterpart, DFrFTs can be interpreted physically as continuous rotations of the associated wave functions through an angle $Z$ in the phase space, see Fig. 1A (18). The idea is thus to construct finite circuits that are capable of imprinting rotations to any light field. To this end, we first note that in quantum mechanics, three-dimensional spatial rotations of complex state vectors are generated via operations of the angular momentum operators $J_k$ $(k = x, y, z)$ on the Hilbert space of the associated system (19). Interestingly, these concepts can be readily translated to the optical domain by mapping the matrix elements of the angular momentum operators over the inter-channel couplings of judiciously engineered waveguide arrays (Fig. 1B) . In doing so, we choose the $J_x$ operator as a prototype in designing our system. In this manner, the photonic lattices are synthesized in such a way that the corresponding coupling matrix mirrors the elements of the $J_x$-matrix: $(J_x)_{m,n} = \kappa_0\big([(j-m)(j+m+1)]^{1/2}\delta_{m+1,n} + [(j+m)(j-m+1)]^{1/2}\ \delta_{m-1,n}\big)/2$. Here, $\kappa_0$ is an arbitrary scale factor and the indices $m$ and $n$ range from $-j$ to $j$ in unit steps. Meanwhile, $j$ represents an arbitrary positive number, integer or half-integer, that determines the total number of waveguides via $N = 2j + 1$ (Fig. 1B).

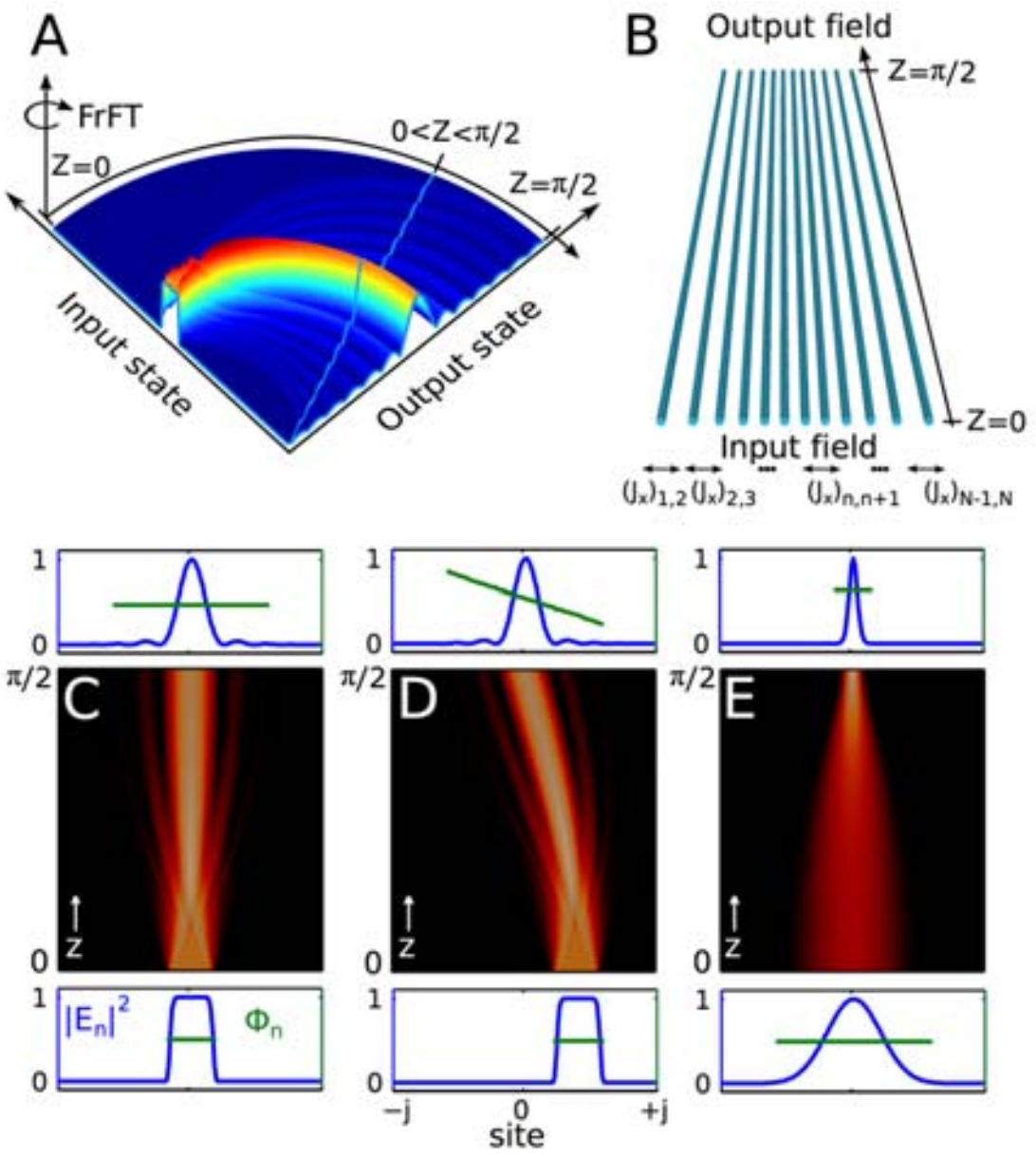

**Fig. 1.** (A) Pictorial view of actual Fractional Fourier transforms exemplified as continuous rotations in phase space. (B) Schematic representation of a pre-engineered $J_x$-lattice. (C-E) Top views of continuous "rotations" of a rectangular (C), displaced rectangular (D), and Gaussian (E) optical wave functions in a $J_x$-lattice with $N = 151$. The bottom and top plots show the intensities of the ingoing and outgoing wave packets, respectively. The green lines describe the phase distributions of the optical fields.

It can be shown that the evolution of light in a $J_x$-lattice is governed by the following set of $N$ equations (20)

$$i\frac{d}{dZ}E_m(Z) = \frac{1}{\kappa_0}\sum_{n=-j}^{j}(J_x)_{m,n}\,E_n(Z). \qquad (1)$$

Here, $E_n(Z)$ denotes the mode field amplitude at site $n$, and $Z$ represents a normalized propagation distance. In the quantum optics regime, single-photons traversing such devices are governed by a similar set of Heisenberg equations that are isomorphic to Eq. (1) except that in this case $E_n(Z)$ is replaced by the photon creation operator $a_n^{\dagger}(Z)$. A spectral decomposition of the $J_x$-matrix yields the eigensolutions

$$u_n^{(m)} = 2^n\left(\frac{(j+n)!(j-n)!}{(j+m)!(j-m)!}\right)^{\frac{1}{2}} P_{j+n}^{(m-n,-m-n)}(0), \qquad (2)$$

which in combination with the eigenvalues, $\beta_m = -j, \ldots, j$, render a closed-form Green function

$$G_{p,q}(Z) = (-i)^{q-p}\sqrt{\frac{(j+p)!\,(j-p)!}{(j+q)!\,(j-q)!}}\left[sin\left(\frac{Z}{2}\right)\right]^{q-p}$$

$$\left[cos\left(\frac{Z}{2}\right)\right]^{-q-p} P_{j+p}^{(q-p,-q-p)}\big(cos(Z)\big). \qquad (3)$$

Note that $q$ and $p$ represent the excited and observed sites, respectively, and $P_n^{(A.B)}(x)$ are the Jacobi polynomials of order $n$ (SI). In our approach, any particular order of the DFrFT arises at one specific propagation distance $Z$ lying between 0 and$\pi/2$. Remarkably, the eigensolutions of the $J_x$-lattices are the actual Hermite-Gauss polynomials sampled at equidistant points separated by distances $h = \sqrt{2/N}$ (7). Evidently, in the limit $N \to \infty$, the $J_x$-eigenfunctions become the Hermite-Gauss polynomials, which in turn are eigenfunctions of the fractional Fourier operator (7). As a result, in the continuous limit, the DFrFT described by Eq. (3) converges to actual FrFT (4, 7); and the standard FT is recovered at $Z = \pi/2$ (Fig. 1C-E). Note that in general the FT obtained in our devices and the usual DFT become equivalent in the continuous limit $N \to \infty$.

For our classical realization of DFrFTs, we use $J_x$-arrays of $N = 21$ waveguides to perform FTs of representative wave packets. The arrays are inscribed into glass chips using femtosecond laser writing technology (22, SI). The value of the scaling factor is tuned to $\kappa_0 = 0.21cm^{-1}$, producing the Fourier plane after a propagation distance of $7.48cm$.

We first consider a Gaussian wave packet with a FWHM covering five sites (Fig. 2A). By means of fluorescence microscopy (23) we monitor the full intensity evolution from the input to the output plane. This fluorescence image (Fig. 2A) shows a gradual transition from an initially narrow Gaussian distribution at the input to a broader one at the Fourier plane (left and right panels Fig. 2A). These observations demonstrate the scaling Fourier theorem – compressions in space correspond to scale expansions in the Fourier space and vice versa. More importantly, these results demonstrate the versatility of our approach in observing and extracting DFrFTs of all orders ($Z \in [0, \pi/2]$) simultaneously. For comparison, we plot the continuous FrFT produced by the corresponding continuous Gaussian profile (red curves Fig. 2A). Apart from the scaling theorem we demonstrate the shift theorem through a translation of the input Gaussian beam by six channels to the edge. This experiment reveals that off-center input fields tend to travel to the center at the output plane (Fig. 2B). Additionally, we can show that our system works for other input configurations (SI).

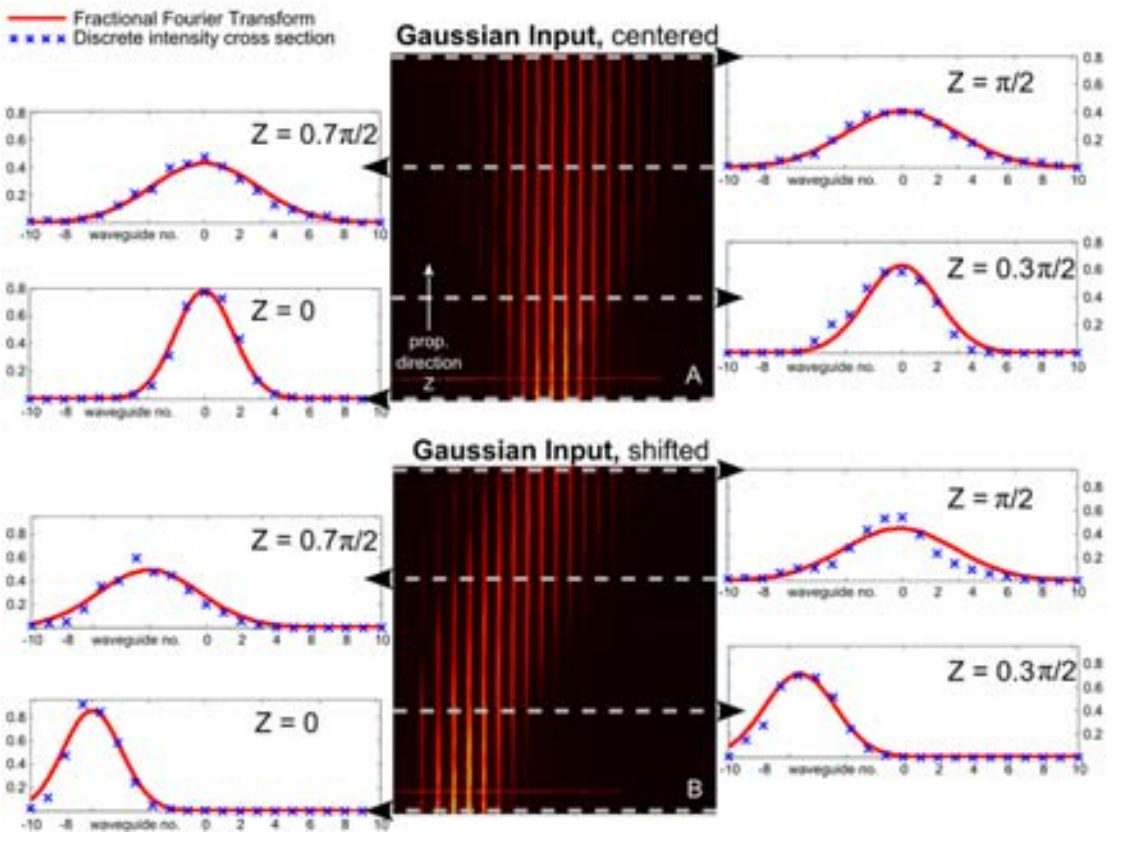

**Fig. 2.** (A) Transformation of a Gaussian input into a Gaussian profile of larger width along the evolution in the $J_x$-lattice. The FT is obtained at $z = \pi/2$. The experimental data (blue crosses) is compared to the analytic FrFT (red curves). (B) A shifted input Gaussian profile evolves towards the center of the lattice and acquires the same width as in (A).

We now turn our attention to the way these $J_x$-lattices respond under single site excitation conditions. Surprisingly, evaluation of the Green function Eq. (3) at the Fourier plane $Z = \pi/2$ reveals that single-site excitations of $J_x$-lattices give rise to field distributions with magnitudes identical to the corresponding eigenstates. In other words, for this discrete system we find a one-to-one correspondence between the excited site number and the eigenstates of the system: excitation of the $q$-th site excites the $q$-th system eigenstate up to well-defined local phases $\left(G_{p,q}\left(\frac{\pi}{2}\right) = (-i)^{q-p} u_p^{(q)}\right)$ (SI). This intriguing effect is shown in the subpanels of Fig. 3A-D along with the theoretical predictions. Due to its finiteness, our system creates a non-uniform amplitude distribution, as predicted by the theory of DFrFT. In the continuous limit the Green functions Eq. (3) tend to the usual FT (7).

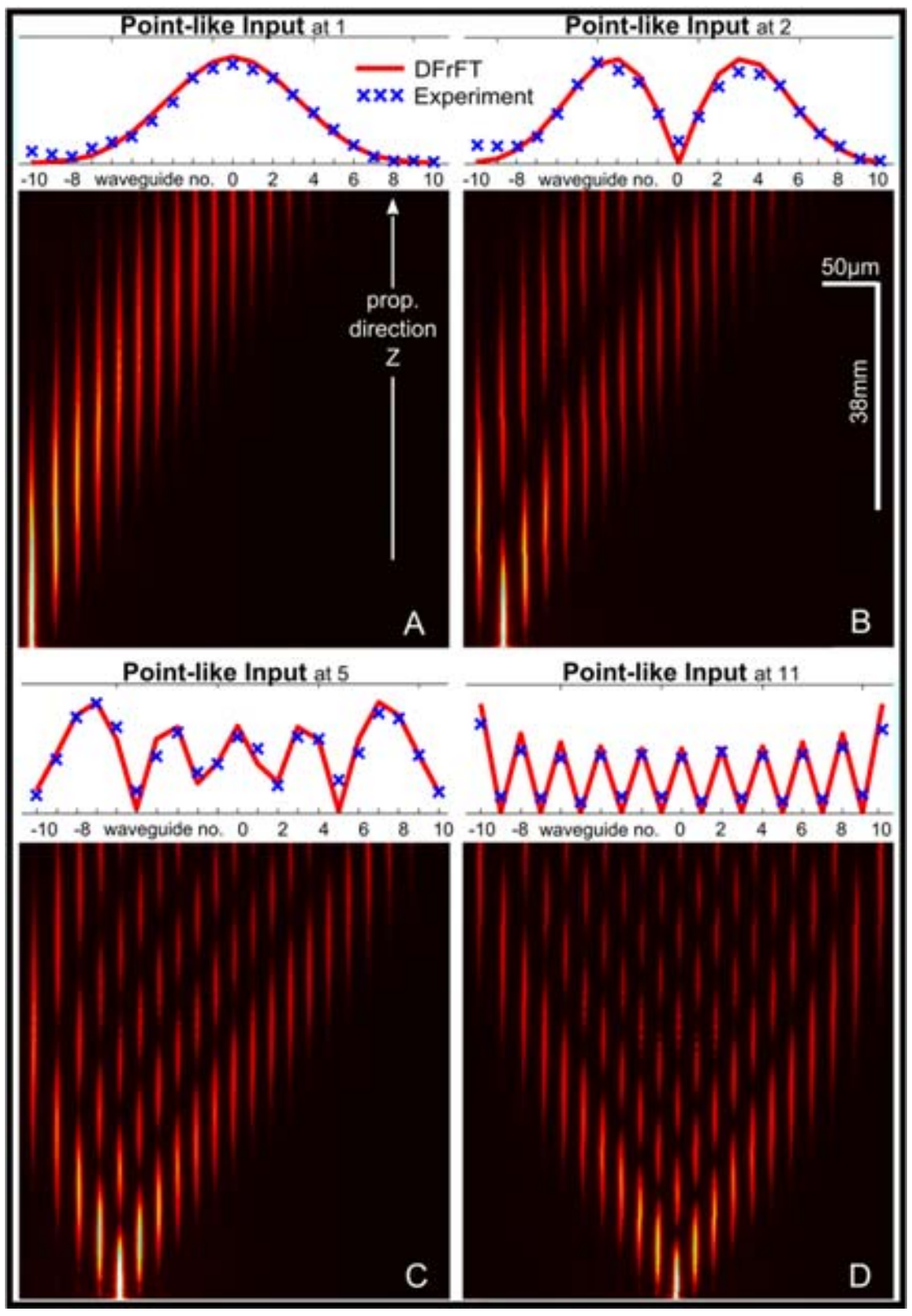

**Fig. 3.** (A-D) Evolution of single-site inputs into the magnitudes of the respective eigensolutions, as predicted theoretically (SI). The experimental data (blue crosses) is compared to the analytic DFrFT (red curves).

At this point it is worth emphasizing the formal equivalence between $J_x$ systems and the so-called quantum Heisenberg XY model in condensed matter physics (24, 25). In this respect, our observations demonstrate the capability of $J_x$ systems in terms of synthesizing any arbitrary eigenstate of XY Hamiltonians by simply exciting specific inputs. To our knowledge, this rather rare property has never been thoroughly investigated before.

In order to demonstrate the applicability of our approach in the quantum domain, we now analyze intensity correlations of separable and path-entangled photon pairs propagating through these Fourier transformers. To do so, we fabricated $J_x$-lattices involving eight elements with a Fourier plane located at $2.62cm$ ($\kappa_0 = 0.6cm^{-1}$). The importance of exploring FTs of such states has been highlighted in several investigations demonstrating interesting effects such as suppression of states and portraying biphoton spatial correlations (26-28).

In this discrete quantum optical context, pure separable two-photon states are readily produced by coupling pairs of indistinguishable photons into two distinct sites $(m, n)$, a state mathematically described by $|\Psi(0)\rangle = a_m^\dagger a_n^\dagger|0\rangle$. Conversely, path-entangled two-photon states are created by simultaneously launching both photons at either site $m$ or $n$ with exactly the same probability, i.e., $|\Psi(0)\rangle = \left[\left(a_m^\dagger\right)^2 + \left(a_n^\dagger\right)^2\right]|0\rangle/2$. Furthermore, the probability of observing one of the photons at site $k$ and its twin at site $l$ is given by the intensity correlation matrix $\Gamma_{k,l}(Z) = \langle a_k^\dagger a_l^\dagger a_l a_k\rangle$ (29). In our experiments the 815-nm photon pairs are produced by a type-I SPDC source.

An intriguing and unique property of the $J_x$ systems is that at $Z = \pi/2$the correlation matrices are given in terms of the eigenstates, as noticed above. Hence, for the separable case, $|\Psi(0)\rangle = a_m^\dagger a_n^\dagger|0\rangle$, the correlation matrices are given by $\Gamma_{k,l} = \left|u_k^{(m)}u_l^{(n)} + u_k^{(n)}u_l^{(m)}\right|^2$, whereas for the entangled state, $|\Psi(0)\rangle = \frac{1}{2}\left[\left(a_m^\dagger\right)^2 + \left(a_n^\dagger\right)^2\right]|0\rangle$, we obtain correlation elements given by $\Gamma_{k,l} = \left|(-i)^{2m}u_k^{(m)}u_l^{(m)} + (-i)^{2n}u_k^{(n)}u_l^{(n)}\right|^2$. Of particular interest is the separable case where the photons are symmetrically coupled into the outermost waveguides, $|\Psi(0)\rangle = a_j^\dagger a_{-j}^\dagger|0\rangle$. In this scenario, only the correlation matrix elements with $(k + l)$ odd are nonzero, and are given by

$$\Gamma_{k,l} = 4^{k+l+1}(j+k)!\,(j-k)!\,(j+l)!\,(j-l)!$$
$$\times\left(\frac{P_{j+k}^{(j-k,-j-k)}(0)P_{j+l}^{(-j-l,j-l)}(0)}{(N-1)!}\right)^2. \qquad (4)$$

These effects are demonstrated for the initial state $|\Psi(0)\rangle = a_{-\frac{7}{2}}^\dagger a_{\frac{7}{2}}^\dagger|0\rangle$ in Fig. 4A, where concentration and absence of probability in the correlation matrix clearly show that some states are completely suppressed – a hallmark of any Fourier unitary process (30).

As a second case, we consider a fully symmetric path-entangled two-photon state of the form $|\Psi(0)\rangle = \left[\left(a_j^\dagger\right)^2 + \left(a_{-j}^\dagger\right)^2\right]|0\rangle/2$. Physically, both photons are entering together into the array at either site $j$ or $-j$ with equal probability (31-33). The correlations are determined by $\Gamma_{k,l} = \left|u_k^{(j)}u_l^{(j)} + u_k^{(-j)}u_l^{(-j)}\right|^2$, from which we infer that the probability of measuring photon coincidences at coordinates $(k, l)$ vanishes at sites where the sum $(k + l)$ is odd. In contrast, at coordinates where $(k + l)$ is even, the correlation function again collapses to the expression given in Eq. (4). This indicates that in this path-entangled case the correlation map appears rotated by 90° with respect to the matrix obtained with separable two-photon states. We performed an experiment to demonstrate these predictions using states of the type $|\Psi(0)\rangle = \frac{1}{2}\left[\left(a_{-7/2}^\dagger\right)^2 + \left(a_{7/2}^\dagger\right)^2\right]|0\rangle$, which were prepared using a $50:50$ directional coupler acting as a beam-splitter (32). The whole experiment is achieved using a single chip containing both the state preparation stage followed by a $J_x$-system, yielding to high interferometric control over the field dynamics (SI). The experimental measurements are presented in Fig. 4B. Similarly, suppression of states occurs as a result of destructive quantum interference. As predicted, a closer look into the correlation pattern reveals that indeed the correlation map appears rotated by 90° with respect to the matrix obtained with separable two-photon states.

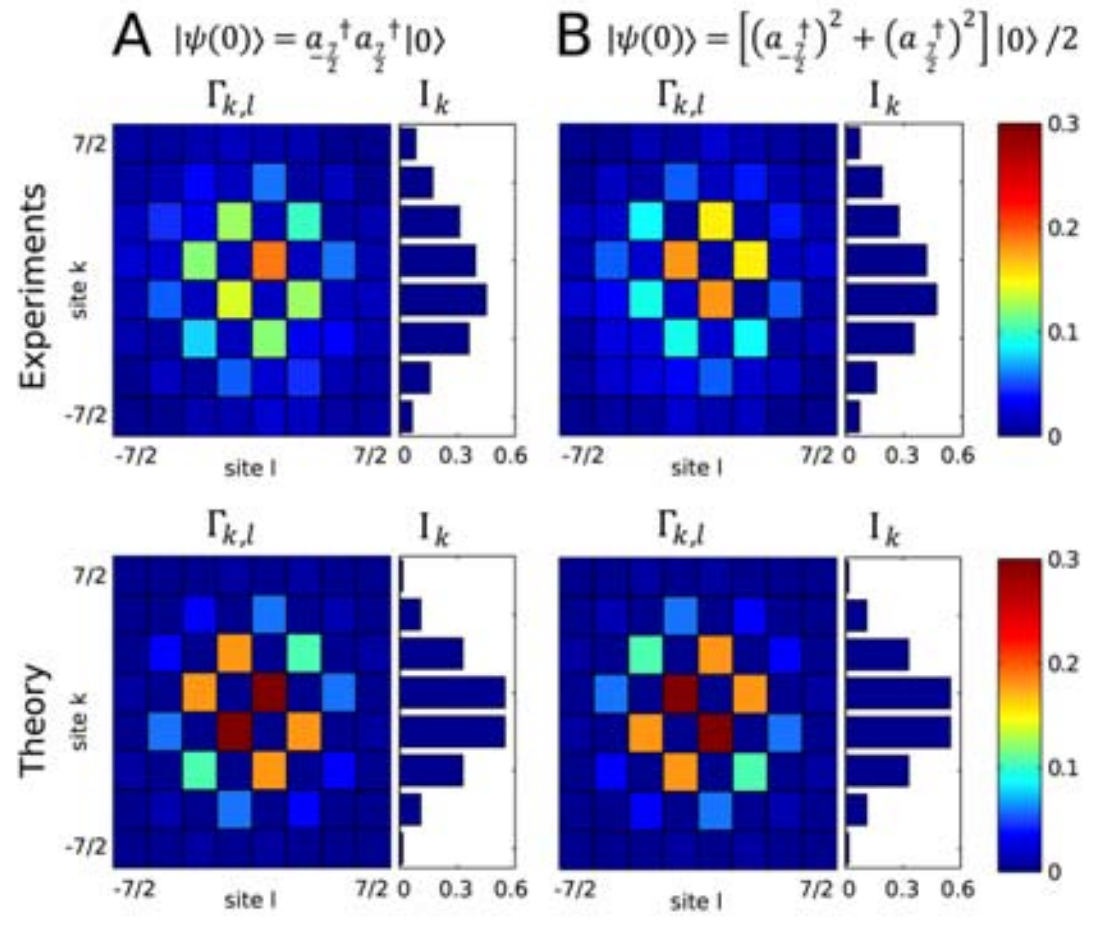

**Fig. 4.** Correlation maps $\Gamma_{k,l}$ of a two-photon state either prepared (A) in a product state or (B) in a path-entangled state after propagating through a $J_x$-lattice. The photon density $I_k$ at the output is shown on the right side of each map.

We emphasize that our measurements feature interference fringes akin to the ones observed in quantum Young's two-slit experiments of biphoton wave functions in free space (26, 27). In those experiments, however, far-field observations were carried out using lenses and the two slits were emulated by optical fibers (26). Along those lines, we have created a fully integrated quantum interferometer to observe fundamental quantum mechanical features. This additionally suggests an effective way to generate quantum states containing only even (odd) non-vanishing interparticle distance probabilities for the separable input state (symmetric path-entangled state).

In conclusion, we have successfully demonstrated a universal discrete optical device capable of performing classical and quantum DFrFTs. Our studies might find applications in developing a more general quantum suppression law (30) and perhaps in the development of new quantum algorithms. Finally, the eigenfunctions associated with the Hamiltonian system explored in our work are specific Jacobi polynomials, which are well known as the optimal basis for quantum phase retrieval algorithms (34).

## SUPPLEMENTARY MATERIAL

This supporting online material contains a theoretical description of the relation between the Fractional Fourier operator and harmonic oscillator systems (S1). In section S2, we show that in the continuous limit $N \to \infty$, the eigenvalue equation describing $J_x$-systems becomes the eigenvalue equation of the quantum harmonic oscillator. In S3 it is shown that at $Z = \pi/2$, the Green function of $J_x$-systems becomes proportional to the amplitude of the corresponding eigenstates. Finally, in S4 we present additional classical experiments and give a description of the setups developed to carry out our quantum and classical experiments.

### S1 The fractional Fourier operator and the quantum harmonic oscillator

In this section, we briefly describe the relation between the continuous fractional Fourier operator and the Hamiltonian of the quantum harmonic oscillator (V. Namias, J. Inst. Maths. Applics **25**, 241-265 (1980)). We start from the fact that the Hermite-Gauss (HG) polynomials of order $n$ are eigenfunctions of the Fourier operator $\hat{F}_Z$ with eigenvalues $\lambda_n = \exp(inZ)$

$$\hat{F}_Z\left\{\exp\left(-\frac{x^2}{2}\right)H_n(x)\right\} =$$

$$\exp(inZ)\exp\left(-\frac{x^2}{2}\right)H_n(x), \qquad \text{(S1)}$$

here, $Z$ is a parameter. Concurrently, from quantum mechanics we know that such HG polynomials are also eigenfunctions of the harmonic oscillator. As a result, it is possible to establish a direct correspondence between the Fourier operator $\hat{F}_Z$ and the Hamiltonian of the quantum harmonic oscillator. To explain this connection, consider the eigenvalue equation

$$\hat{F}_Z\{\Psi_n(x)\} = \exp(inZ)\Psi_n(x), \qquad \text{(S2)}$$

where $\Psi_n(x)$ represents the eigenfunctions of the operator $\hat{F}_Z$ and $\exp(inZ)$ the corresponding eigenvalues. From Eq. (S2) it is clear that a possible representation of $\hat{F}_Z$ is of the form $\hat{F}_Z = exp(iZ\hat{A})$, such that Eq.(S2) becomes

$$exp(iZ\hat{A})\Psi_n(x) = \exp(inZ)\Psi_n(x). \qquad \text{(S3)}$$

Differentiating both sides of Eq. (S3) with respect to $Z$ and evaluating at $Z = 0$, one finds that $\hat{A}$ satisfy the eigenvalue equation

$$\hat{A}\Psi_n(x) = n\Psi_n(x). \qquad \text{(S4)}$$

In order to find the eigenfunctions $\Psi_n(x)$ satisfying Eq.(S4), consider the differential equation

$$\left(-\frac{1}{2}\frac{\mathrm{d}^2}{\mathrm{d}x^2} + x\frac{\mathrm{d}}{\mathrm{d}x}\right)\mathrm{H}_n(x) = nH_n(x), \qquad \text{(S5)}$$

for the Hermite polynomials of order $n$, $H_n(x)$. Then, using the operator identities $1 = \exp\left(\frac{x^2}{2}\right)\exp\left(-\frac{x^2}{2}\right)$, and $\exp\left(\frac{\xi x^2}{2}\right)\left(\frac{\mathrm{d}^n}{\mathrm{d}x^n}\right)\exp\left(-\frac{\xi x^2}{2}\right) = \left(\frac{\mathrm{d}}{\mathrm{d}x} - \xi x\right)^n$, one can show that Eq. (S5) can be written as

$$\left(-\frac{1}{2}\frac{\mathrm{d}^2}{\mathrm{d}x^2} + \frac{1}{2}x^2 - \frac{1}{2}\right)\exp(-x^2/2)\mathrm{H}_n(x),$$

$$= n\exp(-x^2/2)H_n(x) \qquad \text{(S6)}$$

where we have used the commutator $\left[x, \frac{\mathrm{d}}{\mathrm{d}x}\right] = -1$. Comparing Eq. (S4) and Eq. (S6) one can see that $\widehat{\mathrm{A}}$ becomes

$$\widehat{\mathrm{A}} = \left(-\frac{1}{2}\frac{\mathrm{d}^2}{\mathrm{d}x^2} + \frac{1}{2}x^2 - \frac{1}{2}\right). \qquad \text{(S7)}$$

Consequently, the operator $\hat{F}_z$ can be written as

$$\hat{F}_z = \exp\left(\frac{iz}{2}\left(-\frac{\mathrm{d}^2}{\mathrm{d}x^2} + x^2 - 1\right)\right) \qquad \text{(S8)}$$

which indicates that $\hat{F}_z$ is given as the exponential of the Hamiltonian of the harmonic oscillator (V. Namias, J. Inst. Maths. Applics **25**, 241-265 (1980)). This in turn shows that the eigenfunctions $\Psi_n(x)$ satisfying Eq. (S2) are the HG polynomials of order $n$

$$\hat{F}_z\{\exp(-x^2/2)H_n(x)\}_z.$$

$$= \exp(inz)\exp(-x^2/2)H_n(x) \qquad \text{(S9)}$$

Since there is a one-to-one correspondence between the dynamics of the quantum harmonic oscillator and the fractional Fourier operator, the implementation of such transform should be immediate using harmonic oscillator systems (J. Opt. Soc. Am. A **12**, 1448 (1995)).

## S2 $J_x$ photonic lattices as discrete harmonic oscillators

Our aim in this section is to show that in the continuous limit $N \to \infty$, the eigenvalue equation for $J_x$-lattices becomes the eigenvalue equation of the quantum harmonic oscillator. Consider the matrix representation of the $J_x$ operator

$$(J_x)_{m,n} = \frac{1}{2}\Big(\sqrt{j(j+1) - m(m+1)}\delta_{n,m+1} + \sqrt{j(j+1) - m(m-1)}\delta_{n,m-1}\Big) \qquad \text{(S10)}$$

The indices $m$ and $n$ range from $-j$ to $j$ in unit steps and $j$ is an arbitrary positive integer or half-integer. The dimension of the $J_x$ matrix is $N = 2j+1$. We now introduce the variable $\gamma = j(j+1) = (N^2 - 1)/4$, which implies that $(J_x)_{m,n}$ can be written as

$$(J_x)_{m,n} = \sqrt{\gamma}\left(\sqrt{1 - \frac{1}{\gamma} m(m+1)}\delta_{n,m+1} + \sqrt{1 - \frac{1}{\gamma} m(m-1)}\delta_{n,m-1}\right)/2 \qquad \text{(S11)}$$

Let us consider the eigenvalue equation for this matrix

$$\frac{\sqrt{\gamma}}{2}\left(\sqrt{1 - \frac{1}{\gamma} m(m+1)}\psi_{m+1} + \sqrt{1 - \frac{1}{\gamma} m(m-1)}\psi_{m-1}\right) = \beta_m \psi_m \qquad \text{(S12)}$$

Considering the region $m \ll j$, since $\gamma \propto N^2$, in the limit $N \to \infty$, the terms $m(m \pm 1)/\gamma << 1$. Hence, in the domain far from the edge of the array a Taylor expansion yields

$$\sqrt{\gamma}\sqrt{1 - \frac{m}{\gamma}(m \pm 1)} \approx \sqrt{\gamma}\left(1 - \frac{m}{2\gamma}(m \pm 1) - \frac{m^2}{8\gamma^2}(m \pm 1)^2\right) \qquad \text{(S13)}$$

By defining $m = x\gamma^{1/4}$ $\left(\text{or } x = m/\gamma^{1/4}\right)$, we obtain

$$\sqrt{\gamma}\sqrt{1 - \frac{m}{\gamma}(m \pm 1)} \approx \gamma^{1/2} - \frac{x^2}{2} \mp \frac{x}{2\gamma^{1/4}} - \frac{x^4}{8\gamma^{1/2}} \mp \frac{x^3}{4\gamma^{3/4}} - \frac{x^2}{8\gamma}, \qquad \text{(S14)}$$

up to terms containing the factor $\frac{1}{\gamma^{1/2}}$

$$\sqrt{\gamma}\sqrt{1 - \frac{m}{\gamma}(m \pm 1)} \approx \gamma^{1/2} - \frac{x^2}{2} \mp \frac{x}{2\gamma^{1/4}} - \frac{x^4}{8\gamma^{1/2}} \qquad \text{(S15)}$$

Plugging this expression into Eq. (17)

$$\left(\gamma^{1/2} - \frac{x^2}{2} - \frac{x}{2\gamma^{1/4}} - \frac{x^4}{8\gamma^{1/2}}\right)\psi_{m+1} + \left(\gamma^{1/2} - \frac{x^2}{2} + \frac{x}{2\gamma^{1/4}} - \frac{x^4}{8\gamma^{1/2}}\right)\psi_{m-1} = 2\beta_m \psi_m \qquad \text{(S16)}$$

We redefine the functions $\psi_m = \psi(x) = \psi\left(\frac{m}{\gamma^{1/4}}\right)$, and $\psi_{m+1} = \psi\left(x + \frac{1}{\gamma^{1/4}}\right) = \psi\left(\frac{m}{\gamma^{1/4}} + \frac{1}{\gamma^{1/4}}\right)$ such that we can introduce the Taylor series

$$\psi_{m\pm1} = \psi\left(x \pm \frac{1}{\gamma^{1/4}}\right) = \psi(x) \pm \frac{1}{\gamma^{1/4}}\psi'(x) + \frac{1}{2\gamma^{1/2}}\psi''(x), \qquad \text{(S17)}$$

Where again we have kept terms up second order in $1/\gamma^{1/4}$. Substituting Eq. (22) into Eq. (21), and using the limit

$$\lim_{N\to\infty} \Delta = \lim_{N\to\infty} \sqrt[4]{\left(\frac{4}{N^2}\right)\left(\frac{1}{1 - \frac{1}{N^2}}\right)} = 0.$$

We obtain the time-independent Schrödinger equation for the quantum harmonic oscillator

$$\left(-\frac{1}{2}\frac{d^2}{dx^2} + \frac{1}{2}x^2\right)\psi(x) = (\sqrt{\gamma} - \beta)\psi(x). \qquad \text{(S19)}$$

Therefore, in the continuous limit $N \to \infty$, the difference equation describing $J_x$ photonic lattices becomes the time-independent Schrödinger equation for the quantum harmonic oscillator. Note, however, that due to the importance of the condition $m \ll j$ in this derivation this statement is only valid when dealing with signals that are square integrable in the continuous limit. Then, $J_x$ lattices are a discrete version of the quantum harmonic oscillator.

## S3 Green function for $J_x$ photonic lattices

The evolution of light in $J_x$-lattice is governed by the set of $N$ coupled differential equations

$$i\frac{d}{dZ}E_m(Z) = \sum_{n=-j}^{j}(J_x)_{m,n}\,E_n(Z). \qquad \text{(S20)}$$

The normalized propagation coordinate $Z$ is given by $Z = \kappa_0 z$, where $z$ is the actual propagation and $\kappa_0$ is an arbitrary scale factor, and $E_n(Z)$ denotes the mode field amplitude at site $n$. A spectral decomposition of the $J_x$-matrix yields the eigensolutions

$$u_n^{(m)} = (2)^n\sqrt{\frac{(j+n)!(j-n)!}{(j+m)!(j-m)!}}\,P_{j+n}^{(m-n,-m-n)}(0), \quad \text{(S21)}$$

$P_n^{(A.B)}(x)$ are the Jacobi polynomials of order $n$. And the corresponding eigenvalues are integers or half-integers, $\beta_m = -j,\dots,j$, depending on the parity of $N$ (L. M. Narducci and M. Orzag, Am. J. of Phys. **40**, 1811 (1972), A. Perez-Leija, R. Keil, H. Moya-Cessa, A. Szameit, and D. N. Christodoulides, Phys. Rev. A **87**, 022303 (2013)). Using the eigenvectors and eigenvalues we obtain the Green function

$$G_{p,q}(Z) = \sum_{r=-j}^{j} u_q^{(r)}u_p^{(r)}\exp(irZ). \qquad \text{(S23)}$$

$G_{p,q}(Z)$ represents the amplitude at site $p$ after an excitation of site $q$. Using Eq. (S21) and the properties of the Jacobi polynomials one can show that Eq. (S23) reduces to the closed-form expression

$$G_{p,q}(Z) = (-i)^{q-p}\sqrt{\frac{(j+p)!\,(j-p)!}{(j+q)!\,(j-q)!}}\left[\sin\left(\frac{Z}{2}\right)\right]^{q-p}$$

$$\times\left[\cos\left(\frac{Z}{2}\right)\right]^{-q-p} P_{j+p}^{(q-p,-q-p)}(\cos(Z)). \quad \text{(S25)}$$

Evaluation of Eq. (S25) at $Z = \pi/2$ yields

$$G_{p,q}\left(\frac{\pi}{2}\right) = (-i)^{q-p}(2)^p\sqrt{\frac{(j+p)!(j-p)!}{(j+q)!(j-q)!}}\,P_{j+p}^{(q-p,-q-p)}(0)$$

$$= (-i)^{q-p}u_p^{(q)} \qquad \text{(S26)}$$

Eq. (S26) shows that at $Z = \pi/2$ the Green function becomes proportional to the amplitude of the corresponding eigenstates depending on the excited site. In other words, there is a one-to-one correspondence between the excited site number and the eigenstates of the system: excitation of the q-th site excites the p-th eigenstate up to well-defined local phases.

## S4 Experimental setup

### DFrFT of Top-hat functions

As additional classical experiments we examine a well-known example of a FT pair, the top-hat function whose FT is a sinc function. This input state is intentionally chosen to illustrate the difference between the FT and the DFrFT when the continuous limit is not met. The input fields are prepared by tailoring laser light at a wavelength of 632 nm using a SLM. In Fig. S1A, we show the initial top-hat pattern covering 4 waveguides at the edge of the array (top panel). We thus break the correspondence to the harmonic oscillator as explained at the end of section S2. In the bottom panel of Fig. S1A the measured output field is compared to the computed DFrFT. Although the agreement between the measurement in the $J_x$-photonic lattice and the theoretical DFrFT is very good, one can clearly see slight deviations between the DFrFT and a sinc function. Furthermore, we launch a centered broader top-hat distribution with a linear phase ramp (Fig. S1B). This is achieved by tilting the array with an angle of approximately 0.01°. As can be seen, the ramping phase in the input field does not result in a pure spatial displacement at $Z = \pi/2$. Again there are deviations between the DFrFT and the shifted sinc function.

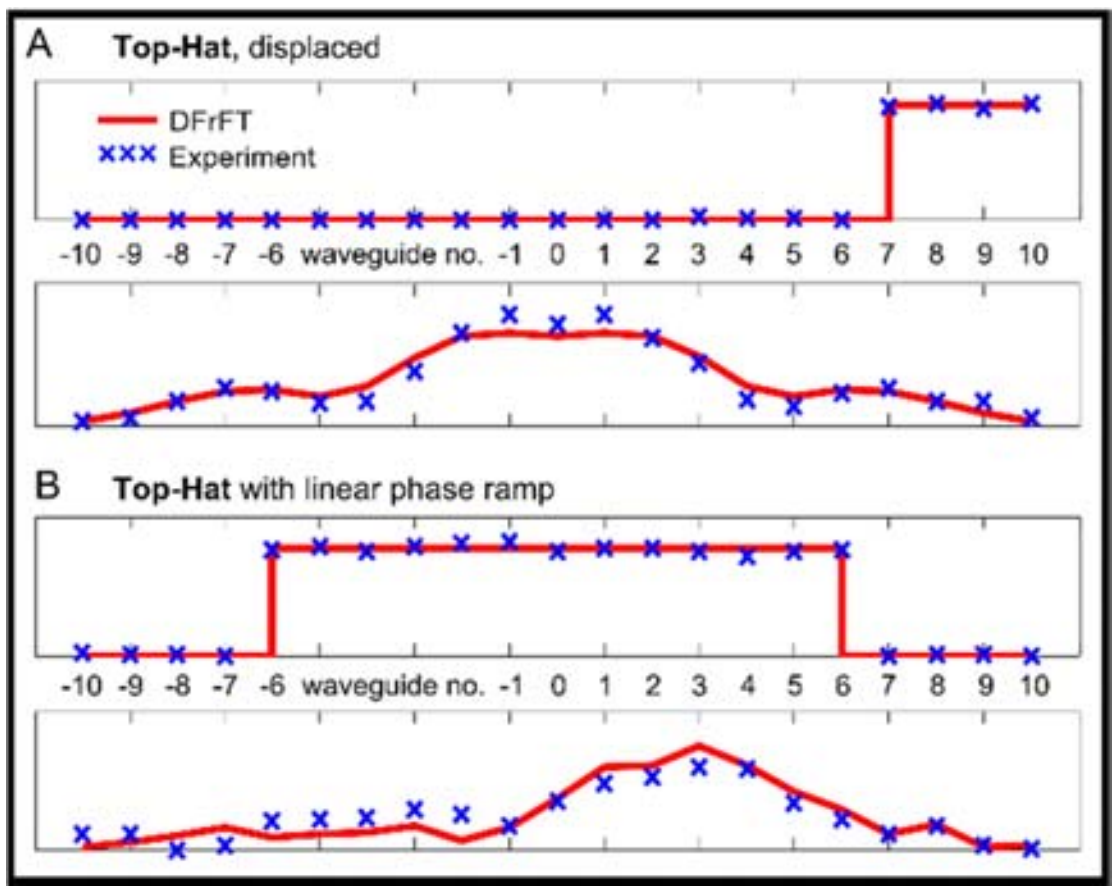

**Figure S1:** Experimental observation of DFrFTs of top-hat functions launched (A) at the edge and (B) at the center (with a linear phase ramp) of a $J_x$-photonic lattice. Upper panels in A and B show the respective input intensity distributions while the lower panels show the output intensity distributions.

### Devices fabrication

Our devices are fabricated in bulk fused silica samples (Corning 7980, ArF grade) using the femtosecond laser direct-write approach (A. Szameit and S. Nolte, J. Phys. B: At. Mol. Opt. Phys. **43**, 163001 (2010)). The transparent material is modified within the focal region, which results in a local increase of the refractive index.
For the samples illuminated with single-photon states of light, we used a RegA 9000 seeded by a Mira Ti:Sa femtosecond laser oscillator. The amplifier produced 150 fs pulses centered at 800 nm at a repetition rate of 100 kHz, with energy of 450 nJ. The structures were permanently inscribed with a 20x objective while moving the sample at a constant speed of 60 mm/min by high-precision positioning stages (ALS 130, Aerotech Inc.). The mode field diameters of the guided mode were 18 μm × 20 μm at 815 nm. All structures were designed with fan-in and fan-out sections arranged in a three-dimensional geometry and located prior and after the $J_x$-lattice, respectively. This effectively suppresses any unwanted crosstalk between the guides and permits easy coupling to fiber arrays with a standard spacing of 127 μm.

For the fabrication of the devices used with classical light, we employed an Yb-doped fiber laser (Amplitude Systèmes) operating at a wavelength of 532 nm, a repetition rate of 200 kHz and a pulse length of 300 fs. Waveguides were written with 300 nJ pulses focused by a 20x objective. The sample was moved at a velocity of 200 mm/min using high-precision positioning stages. The mode field diameters of the guided mode were 10 μm × 10 μm at 632 nm.

**Amplitude and phase modulation of classical light**

In order to be able to shape the input field at will, we employ a SLM (Holoeye Pluto VIS) for amplitude and phase modulation. An initial beam at a wavelength of 632 nm is expanded to homogeneously illuminate the display of the SLM. The reflected beam is Fourier transformed by a 300 mm spherical lens and spatially filtered by a slit aperture. The so prepared field distribution is scaled down to micrometer size by means of a 4f configuration involving a 250 mm lens and a 20x microscope objective (Olympus Plan Achromat). For the sake of maximizing the efficiency of coupling the free space electric field distribution into the discrete array of single-mode waveguides, the beams are prepared in arrays of narrow Gaussian spots matching the mode profile of the waveguides. Amplitudes and positions of all individual Gaussian spots are adjusted according to the desired input beam profile and the positions of the waveguides in the input plane.

**Experimental setup for characterization of two-photon correlations**

A $BiB_3O_6$ nonlinear crystal was pumped with a 70 mW CW pump laser emitting at 407.5 nm, which provided pairs of indistinguishable photons due to type I spontaneous parametric down-conversion (SPDC). Photon pairs with a central wavelength of 815 nm were filtered by 3-nm FWHM interference filters. They were further coupled to the chip via fiber arrays, and subsequently fed into single-photon detectors (avalanche photodiodes). The two-photon correlation function was determined by analyzing the two-fold coincidences recorded between all output channels with the help of an electronic correlator card (Becker & Hickl: DPC230). The coincidences were analyzed with a time window set at 5 ns and are corrected for detector efficiencies. Accidental coincidences due to simultaneous detection of two photons not coming from the same pair are estimated to occur with a negligible rate of less than $2 \times 10^{-6}$ per second. Non-deterministic number-resolved photon detection was achieved using fiber beam-splitters.